# "Half – Bogoliubons" as the intermediate states for the phase coherence in underdoped cuprates

Han Li, Zhaohui Wang, Shengtai Fan, Jiaseng Xu, Huan Yang✉, and Hai-Hu Wen✉

Center for Superconducting Physics and Materials, National Laboratory of Solid State Microstructures and Department of Physics, Collaborative Innovation Center for Advanced Microstructures, Nanjing University, Nanjing 210093, China

✉e-mail: huanyang@nju.edu.cn; hhwen@nju.edu.cn

## Abstract

**Superconductivity is achieved by the pairing of electrons and phase coherence between the Cooper pairs. According to the Bardeen-Cooper-Schrieffer theory, the quasiparticles with Bogoliubov dispersion exists and reveal particle-hole symmetric coherence peaks on the single particle tunneling spectrum. Here we report the observation of two kinds of tunneling spectra showing only one side of the 'coherence peak' but with symmetric energies (about ±11 meV) in underdoped cuprate superconductor $Bi_2Sr_{2-x}La_xCuO_6$ single crystals ($p \approx 0.114$) with fractional superconductivity. Merging these two kinds of spectra can mimic the complete Bogoliubov dispersion, thus we name the electronic states associated with these half-peaked spectrum as 'half Bogoliubons'. In previous studies, it was shown that two doped holes may bind into a local pair within the $4a_0 \times 4a_0$ plaquette of CuO bonds[1,2] ($a_0$: distance between nearest Cu atoms). We attribute the 'half-Bogoliubons' to the intermediate states for the phase coherence, and they**

**correspond to the three-hole and one-hole states as the excited ones from the local pairing state of two holes. An entanglement of these two 'half-Bogoliubons' would mean the dynamic hopping of charge freedom resulting in the phase coherence between the local paired states. Our results unravel a unique process for establishing the phase coherence through exchanging a charge between the regions with local pairs.**

## Main Text

### Introduction

Superconductivity occurs when electrons pair up to form Cooper pairs and then condense into the superconducting condensate through phase coherence. The superconducting state is protected by the superconducting gap $\Delta$, and the superfluid density governs the phase stiffness. According to the Bardeen-Cooper-Schrieffer (BCS) theory[3], the elementary excitations of the superconducting ground state are Bogoliubov quasiparticles, which is the quantum-mechanical particle-hole mixture[4]. The annihilation operator of excited Bogoliubov quasiparticle (Bogoliubon) is a linear combination of electron and hole operators with the amplitudes of $u_k$ and $v_k$ via $\gamma_{k\uparrow} = u_k c_{k\uparrow} + v_k c^\dagger_{-k\downarrow}$, and obey the constraint for the coherence factors $|u_k|^2 + |v_k|^2 = 1$ for any momentum $k$ (normalization)[4-6]. In BCS formalism, the Bogoliubov dispersion forms two branches separately at positive (+) and negative (–) energies $E_k^\pm = \pm\sqrt{\xi_k^2 + |\Delta_k|^2}$, where $\xi_k$ is the quasiparticle energy counted from the Fermi energy ($E_F$)[7]. The Bogoliubov band back-bending results in two symmetric coherence peaks at the gap energies $\pm\Delta$, and the BCS coherence factors are closely related to the spectral weight ($|v_k|^2$ below $E_F$ and $|u_k|^2$ above)[8]. The coherence peaks of Bogoliubov quasiparticles and superconducting gapped features are the natural consequence of the superconducting state[9,10].

In cuprates, the superconductivity originates from hole doping to the Mott insulators[11-13]. One peculiar point when compared to BCS superconductors is that the electron pairing and phase coherence in cuprates do not happen simultaneously[14-16]. However, what is sure is that the charge carriers in the superconducting state of cuprates are still the condensed "bosons" that carry twice of electron charges[17], just like the Cooper pairs of BCS superconductors. Meanwhile, spectroscopy evidence has demonstrated the BCS-like particle-hole symmetric spectrum in cuprates[4,8,18] besides a background due to the strong correlation effect[19]. This is especially true as evidenced by the spectra measured at the optimal doping. Thus, it has been confirmed that the elementary excitations in the superconducting state are still the Bogoliubov quasiparticles (we term them as Bogoliubons) like that in BCS superconductors[10,20,21]. However, because of the existence of pseudogaps[12] at high energies and correlation effect, this symmetric feature may be strongly distorted[19]. In the extremely underdoped region, although superconductivity can still be detected by resistivity or magnetization measurements, the coherence peaks are almost invisible, showing a very weak coherence weight of quasiparticles.

Another important issue in cupartes is the existence of the $4a_0 \times 4a_0$ plaquette with internal stripes in the pseudogap phase[22,23], and this order can also emerge in the region of the vortex cores[24,25]. Recent experiments provide evidence that the pseudogap phase contains a $8a_0$-period pair density wave (PDW) state[26] and the $4a_0$-period plaquette was attributed to the secondary charge density modulation of such PDW[25,27]. These spectroscopic evidence strongly indicate the preformed incoherent Cooper pairs in the pseudogap phase[16,28,29]. Moreover, the spatial occupation of the $4a_0 \times 4a_0$ plaquettes systematically evolves with increasing hole doping[2], and statistics tells that each $4a_0 \times 4a_0$ plaquette contains approximately two holes on average, which is believed to be closely related to the preformed local Cooper pairing[1]. Consistently, recent investigation suggests a 'pairing glue' originates from the quantum phase strings created by the hopping of holes in the spin background[30]. An electronic state with a $4a_0 \times 4a_0$ plaquette structure was discovered with approximately two holes within each plaquette, and it was proposed that local pairing may be formed through this two-hole bound states[1,2]. This "two-hole" ground state ansatz[30] may capture the significant ingredients of the pairing force in cuprate superconductors. The key mystery now is to understand how the phase coherence is established between the preformed pairs within neighboring $4a_0 \times 4a_0$ plaquettes in order to have the superconducting state with a macroscopic phase coherence.

In this work, we report the observation of two kinds of asymmetric tunneling spectra showing only one 'coherence peak' either at positive or negative energies in some regions of the underdoped $Bi_2Sr_{2-x}La_xCuO_{6+\delta}$ ($x = 0.75$, La-Bi2201) by using scanning tunneling microscopy or spectroscopy (STM/STS). Interestingly, merging these two kinds of spectra mimic that of a complete Bogoliubov dispersion, we thus name these one-side-peaked spectra as a 'half-Bogoliubon'. It is found that the so-called 'half-Bogoliubon' peak has a strong temperature dependence, but is not suppressed by a magnetic field up to 9 Tesla applied perpendicular to $ab$-plane. Based on the picture concerning the ground state of 'two-hole bound state in one $4a_0 \times 4a_0$ plaquette', and we attribute the 'half-Bogoliubons' as the intermediate states for the phase coherence between the local paired states.

## Results

## Characterization of La-Bi2201 sample

Figure 1a shows the schematic phase diagram of the La-Bi2201 system which holds the AF order until the nominal hole density per Cu reaches $p \sim 0.10$. Superconductivity emerges with further doping, and the maximum critical temperature $T_c$ is about 35 K at the hole doping level of $p \sim 0.16$[31]. The La-

Bi2201 sample with a nominal La doping level of $x = 0.75$ studied in this work corresponds to the hole doping level of $p \sim 0.114$[32], which is higher than the threshold for the insulator-superconductor transition ($p \sim 0.10$). Figures 1b and 1c display the magnetic and transport characterizations of the La-Bi2201 samples, respectively. Both curves show the bulk zero resistance superconducting transition at about 15 K. From the broad magnetic transition, we conclude that the superconductivity is still fractional and inhomogeneous. Figure 1d shows a typical atomically-resolved topography of the BiO cleavage surface, and the super-modulation can be clearly observed on the surface. Figure 1e shows a set of tunneling spectra in a wide bias-voltage range recorded at points along the arrowed line in Fig. 1d. The spectra look generally uniform, indicating that the electronic structure is almost uniform in a large energy range. When the measurement is conducted within a small energy window, the spectra will become inhomogeneous, as shown below. In the extremely undoped sample with $p \sim 0.08$, a charge transfer gap was observed with the gap edge at about −0.2 and +1.5 eV; and the gap is filled up with partial density of states (DOS) with further doping, see for example in the sample with $p \sim 0.10$[1]. The spectra measured in present sample follow the trace of the filled charge transfer gap and the asymmetric high-energy differential conductance respect to $E_F$ as observed in the sample with $p \sim 0.10$[1]. The asymmetry of high-energy conductance is consistent with previous report on lightly hole-doped $Ca_{2-x}Na_xCuO_2Cl_2$ (Na-CCOC)[22], and it reveals a larger probability for electron extraction than for injection, as anticipated for a doped Mott insulator[19, 33]. In the low-bias range, an energy gap can be observed at energies within ±100 meV in Fig. 1e. This gap is also observed in La-Bi2201 samples with lower doping levels[1], and it is regarded as a large pseudogap or the so-called antiferromagnetic (AF) pseudogap when the charge-transfer gap[34] is completely filled. It should be noted that the coherence peaks associated with the superconducting gap is not visible here because of the very small superconducting gap, very diluted superfluid density and the very wide bias-voltage range.

## Tunneling spectra with 'half-Bogoliubon'

In the La-Bi2201 sample with $p \sim 0.10$, accompanied by the filling of the charge transfer gap, we observed the electronic structure with $4a_0 \times 4a_0$ basic plaquettes, and meanwhile, there are three unidirectional bars in each plaquette[1]. It is interesting to see whether such a peculiar electronic structure exists in the present sample with a higher doping level of $p \sim 0.114$. Figure 2a shows the differential conductance mapping measured at +15 mV. One can clearly see the $4a_0 \times 4a_0$ plaquettes with three unidirectional bars in one plaquette, following the similar electronic structure observed in the sample with p $\sim 0.10$[1]. The plaquettes and the unidirectional bars can be distinguished through the auto-correlation analysis (Supplementary Fig. 2). The density of the plaquettes corresponds well to the hole doping level if we assume two holes occupying one plaquette as we did in previous work[1]; the density increases obviously in the present sample with a higher doping level and the plaquettes almost cover the whole field of view. The calculated value from the density of the plaquettes is also close to the nominal doping level. The electronic structure is similar to the checkerboard order observed in Na-CCOC[21,22]. The differential conductance mapping at −15 mV is shown in Fig. 2b, and one can also see the plaquettes and the unidirectional bars. However, the outlines of these $4a_0 \times 4a_0$ plaquettes at −15 mV become more blurred than those at +15 mV on average. In both cases, one can see that some plaquettes become overlapped showing some regions with extended stripy bars or more than three stripy bars exist on one patch.

In the differential conductance mapping, beside the $4a_0 \times 4a_0$ plaquettes, we also find some short-range stripe patterns. As an example, two of them are shown within the white dashed square in Figs.

2a and 2b. The stripe patterns consist of several unidirectional bars with an interval of about $4a_0/3$, and the lengths of the patterns are a bit larger than $4a_0$. For example, the bright stripes near spot No. 1 show up at negative biases, while those near spot No. 2 show up at positive biases. However, these stripes show particle-hole asymmetry. One can see that the brightness of the stripes near spot No. 1 at negative biases becomes darken at positive biases, while the ones near spot No. 2 at positive biases is darken at negative biases. The particle-hole asymmetry also appears on the tunneling spectra measured at these points, and the examples are shown in Fig. 2c. One can see a clear peak at about −11 mV on the spectrum measured at spot No. 1 or at about +11 mV on the spectrum measured at spot No. 2, but the peak is absent at about +11 mV on the spectrum measured at spot No. 1 or at about −11 mV on the spectrum measured at spot No. 2. This feature is consistent with the conductance mappings at ±11 mV shown in Figs. 2a and 2b respectively. This indicates that some spectra show only one-side "coherence peak", leaving a small kink on the opposite bias energy. In regions with moderate brightness, for example, at spot No. 3 in Figs. 2a and 2b, one can see only some kinked features on spectra at about ±11 mV, showing a symmetric feature, these regions may correspond to the ones with very fragile superconductivity and thus the coherence peaks are very shallow. We will show data later to support this conclusion. Interestingly, the energies corresponding to the kinks in the spectrum measured at spot No. 3 are roughly consistent with the one-side "coherence" peak of the spectrum measured at spot No. 1 and the negative-bias side of the spectrum measured at spot No. 2. Meanwhile, when we merge the negative-bias side of the spectrum measured at spot No. 1 and the positive-bias side of the spectrum measured at spot No. 2 together, the constructed spectrum is shown in Fig. 2d and it behaves as the typical spectrum of a complete Bogoliubov dispersion for a perfect superconductor[35], i.e., there are two sharp coherence peaks at symmetric energies of positive and negative sides. *This is*

*the key finding of our present work.*

Based on the BCS theory, when the superconducting gap opens in a band, the Bogoliubov dispersion forms two branches at positive and negative energies $E_{\bm{k}}^{\pm} = \pm\sqrt{\xi_{\bm{k}}^2 + |\Delta_{\bm{k}}|^2}$ (Fig. 2e). The Bogoliubov band structure results in the coherence peaks in the energy-dependent DOS at the gap energies (Fig. 2f). The amplitudes of the coherence peaks qualitatively reflect the phase coherence stiffness of condensed Cooper pairs. In the BCS framework, the DOS of coherence peaks at positive energy can be simply quantified by $|u_k|^2$, while that at negative energies is evaluated by $|v_k|^2$. In BCS superconductors, the low energy electronic structure behaves $\text{DOS}(+E) = \text{DOS}(-E)$ because of particle-hole symmetric Bogoliubov excitation $|u_k|^2(+E) = |v_k|^2(-E)$. Therefore, the observed extremely particle-hole asymmetric spectra shown in Fig. 2c do break the common understanding of superconducting excitation. $|u_k|^2 + |v_k|^2 = 1$

## Local particle-hole asymmetry of the unidirectional charge stripes

In order to clearly visualize the detailed structures of unidirectional stripes, we enlarge the region within the dashed square in Fig. 2a,b and show them in Fig. 3a,b. The structural and spatial layout of positive and negative stripes are visually demonstrated, and they are along the directions of the Cu-O-Cu chains. By doing autocorrelation analysis (see Supplementary Fig. 2), we verify that the spatial spacing of modulation for these two kinds of stripes is nearly $4a_0/3$. This spacing is consistent with the stripe distance of the nematic electronic structure within a $4a_0 \times 4a_0$ plaquette in the sample with $p \sim 0.10$[1]. Although most of the plaquettes have the size of $4a_0 \times 4a_0$ in present sample, the local stripes shown in Fig. 3a,b exceed the length of $4a_0$. By taking a comparison of the d$I$/d$V$ mappings measured at the positive and negative biases, the stripes near the arrowed line C1 are very bright in the mapping measured at the positive bias but are very weak in the mapping at negative energies. Consequently, the spectra measured along C1 in Fig. 3c show clear coherence peaks at about +11 mV, but the coherence peaks disappear on the negative bias side, leaving only the kinked feature at about −11 mV. In contrast, the stripes along the arrowed line C2 are very bright in the d$I$/d$V$ mappings measured at the negative bias but vague at positive bias. Meanwhile, the coherence peaks appear only at about −11 mV but unclear at about +11 mV on the spectra measured along C2 as shown in Fig. 3d. At the positions where the mapping intensity is moderate at both positive and negative bias, for example at green dots near the stripes shown in Fig. 2a,b, the spectra do not show coherence peaks but only kinked feature at about ±10 mV. It should be noted that the superconducting gapped feature always exists even when the coherence peaks are absent on the spectra or the half spectra, and the gapped feature is clearer in the corresponding −d$^3I$/d$V^3$ curves shown in Supplementary Fig. 3. Except for the most salient features, spectra in Fig. 3c-e also show some humps at higher energies beyond the superconducting gap, and they may correspond to the pseudogaps at about 20 to 50 meV.

For a well-defined superconducting tunneling spectra, the gap magnitude $\Delta_k$ characterizes the pairing strength while the sharpness of the coherence peaks indicates the phase coherence stiffness. The coherent spectral weight (weight of the coherent peak) can be suppressed by increasing temperature[16,36,37] or external magnetic field[38]. To examine the thermal effect and magnetic field influence on the asymmetric "coherence" peaks in this work, we provide the temperature and magnetic field dependence of the spectra with the so-called 'half-Bogoliubons'. In Fig. 4a-c, we show the temperature evolution for three kinds of typical spectra as displayed in Fig. 2 and Fig.3. One can see that the peaks of 'half-Bogoliubons' at about +10 and −10mV are both significantly suppressed by increasing the temperature from 0.3 K to 20 K. Moreover, the particle-hole symmetric gap which shows a pair of moderate kinks on the spectra at about ±10 mV is filled up by DOS with increasing temperature. Therefore, the particle-hole symmetric gap with moderate kinks reasonably reflects the superconducting pairing energy scale. Then, we measured the spectra by applying magnetic fields up to 9 T with direction perpendicular to the $CuO_2$ plane. This field roughly equals to the upper critical field ($H_{c2}$) as previously reported for La-Bi2201 samples at hole concentration $p$ = 0.114[32]. Figure 4 d,e shows tunneling spectra measured under H = 0 T and 9 T, and neither positive nor negative peaks are suppressed by the magnetic field. This magnetic field insensitive nature distinguishes the sharp particle-hole asymmetric peaks from the superconducting coherence peaks. Then we can naturally raise the questions: What is the intrinsic nature of these states? and What is their relationship with superconductivity?

**'Half – Bogoliubons' and the phase coherence**

To answer these questions, we must consider what happens to the cuprates as it is doped with holes, especially when the hole concentration is close to the 1/8 doping. In the parent phase, the strong on-site Coulomb repulsion $U$ together with $p$-$d$ hybridization results in a charge transfer Mott insulator, and the inter-copper super-exchange interaction $J$ stabilizes a robust spin-half AF order[11,13,39]. When holes with modest density are doped into the $CuO_2$ planes, the spectral weight transfers from high energy scale to fill up the charge transfer gap forming a prominent pseudogap at about 200 meV[1,12,40,41]. The competition between the kinetic energy, spin interaction and Coulomb repulsion makes the pseudogap phase break the $C_4$ symmetry within $CuO_2$ unit cells leading to a locally stripy electronic structure[42-45]. The onset doping level for the emergence of the pseudogap is coincident with the appearance of the $4a_0$-period plaquettes which persist from the region with AF phase until global superconductivity is established[1,2,22]. In Supplementary Fig. 4, we estimate the densities of the $4a_0$ ×

$4a_0$ plaquettes at different hole doping and find that the spatial occupancy of the $4a_0 \times 4a_0$ plaquettes increases with enhanced hole concentration. After the spatial occupation of plaquettes exceeds a threshold corresponding to the transition from insulator to superconductor, the particle-hole symmetric *d*-wave superconducting gap gradually emerges. According to the single particle tunneling mechanism, the positive (negative) bias side of the spectrum is dictated by leaving (attracting) electrons into (from) the system. Thus, if there is a coherent unoccupied hole DOS at a location, it would be able to observe a coherence peak on the positive bias voltage. On the contrary, if locally we have a coherent occupied DOS of electrons, it would allow us to see a coherent peak at the negative bias when the electrons are taken away. This is exactly what we have witnessed here in present experiment. The studied La-Bi2201 sample has a doping level of $p = 0.114$ and it is close to 1/8 hole per $CuO_2$, as analyzed in Supplementary Fig. 4, each $4a_0 \times 4a_0$ plaquette contains approximately two holes. In view that the merged spectrum shown in Fig. 2c,d mimic that of the complete Bogoliubov dispersion for a *d*-wave superconductor, it is reasonable to attribute the spectrum with one-sided coherence peak in Fig. 2c at about −11 meV and +11 meV to the half branches of the Bogoliubov dispersion. The half branch of the Bogoliubov dispersions $E_k^\pm = \pm\sqrt{\xi_k^2 + |\Delta_k|^2}$ (Fig. 2(c) and (d)) can be generally understood as two exclusive excited states of the ground state with approximately two holes in each $4a_0 \times 4a_0$ plaquette. Since the plaquette of carriers may reflect electron correlation nature and play a crucial role in local pairing[46], it is reasonable that the extremely particle-hole asymmetric electronic states detected on the background of $4a_0$-period plaquettes may provide clues for the superconducting phase coherence

## Discussion

Following this line, our observation of extremely particle-hole asymmetric ±10 meV peaks can be well understood as the excited states based on a 'two-hole' ground state wavefunction[30,47]. This scenario is distinct from the usual pairing mechanism like exchanging AF spin fluctuations[48-50] or the RVB pairing[51-53], as the phase strings create spin-current pattern mediating the pairing force for twisted 'two-hole' ground state[30]. Since previous studies show that the Bose condensate of these two-hole pair state could be lead to the superconducting state[47, 54], thus we consider our case by starting from the spatially uniform ground state that two doped holes bound into a local pair within each $4a_0 \times 4a_0$ plaquette of CuO bonds (illustrated as the upper panel in Fig. 4f). With phase coherence to certain extent, the spectra of this ground state may correspond to that shown in Fig. 3e, the kink features at ± 10 meV indicate the energy scale for pairing while the absence of sharp coherence peaks reflect poor phase coherence. We argue that the 'half-Bogoliubon' peaks at −11meV as shown in Fig. 3d could be a single hole excited state from the paired 'two-hole' ground state. This excited state breaks the hole pair within a plaquette, and with an excited hole hopping into nearby plaquette with only one hole left within a $4a_0$ plaquette (illustrated as the lower panel in Fig. 4f). At this location, due to the inhomogeneous hole distribution, there are less holes compared with the two-holes/plaquette by average, thus it is easier to extract electrons away, leading to the particle-hole asymmetric coherence peak at negative sample bias (Fig. 3d). Similarly, at the locations where the holes are slightly rich, we should observe the spectra with a "coherence peak" at positive bias, as shown by Fig. 3c.

According to our interpretations above, the 'half-Bogoliubon' peaks at about ±11mV should correspond to the half branch of the Bogoliubov dispersions $E_k^\pm = \pm\sqrt{\xi_k^2 + |\Delta_k|^2}$ (as illustrated in Fig. 2e,f). The clear temperature dependence of the one-sided "coherence peaks" shown in Fig. 4a,b suggests that these excited states are inherent to superconductivity, although they are not directly the superconducting coherence peaks yet. It is then natural for us to propose that the 'half-Bogoliubon' peaks reflect the two exclusive excited states of the ground state with approximately two holes in each $4a_0 \times 4a_0$ plaquette. As illustrated in the lower panel of Fig. 4f, we attribute the +11 meV peaks as the hole addition bound state for more than two entangled holes within a $4a_0 \times 4a_0$ plaquette. In this case, the additional hole disturbs the two-hole pairing channel, as a result, the neutral Bogoliubov quasiparticles ($\gamma_{k\uparrow} = u_k c_{k\uparrow} + v_k c^{\dagger}_{-k\downarrow}$) collapse to purely hole-like states ($|u_k|^2 \sim 1$), namely the branch of the Bogoliubov dispersion with peaks at +11 meV. In a similar way, a single hole occupied $4a_0 \times 4a_0$ plaquette is equivalent to adding one additional electron into two-hole ground state. As a cartoon picture, the particle-hole asymmetric excitations as 'half-Bogoliubons' manifest as the intermediate states with for phase coherence, showing a new type of quasiparticles in the superconducting state with particle-hole dualism based on bound state of two-hole per plaquette. These exciting states above 'two-hole' ground state should be taken as evidence that the system is approaching a strong pairing situation near 1/8 doping, that is referred to the state with non-overlapping charge $2e$ bosons before global phase coherence. On this basis, the particle-hole asymmetric 'half-Bogoliubons' could be the intermediate states for the phase coherence between local hole pairs. Thus, exchanging holes between the regions with the local pairing in a form of two holes per plaquette would mean a dynamic hopping of the charge freedom between these pairing blocks. This is right the process for building up the phase coherence between the local pairing. It is the spatial inhomogeneity of the hole density that allow some of these intermediate states for the phase coherence be pinned done. By the way, this kind of intermediate states for the phase coherence cannot happen in a BCS superconductor in which the pairing size is large with enormous Cooper pairs entangled each other, thus the pairing and phase coherence occur simultaneously. In the cuprates, the local pairing has been well illustrated in many experiments, the phase coherence can only be established by exchanging the charge freedom. Therefore, we believe the features of 'half-Bogoliubons' reflect just how the phase coherence is established between local hole pairs, which gives rise to superconducting gaps with sharp particle-hole symmetric coherence peaks (as illustrated in Fig. 2f).

In summary, by conducting scanning tunneling spectroscopy measurements in the underdoped regime of the La-Bi2201 high-$T_c$ superconductor, we reveal the extremely asymmetric spectra with one-side "coherence peaks". Based on the local pairing picture of two-hole bound states, we interpret these asymmetric spectra with one coherence peak at positive or negative bias as the intermediate states for the phase coherence between the local pairing blocks, thus they are termed as the "half-Bogoliubons". This is a unique process for establishing the phase coherence in cuprates with preformed pairs. An entanglement of hole-like and electron-like 'half-Bogoliubons' would mean the dynamic hopping of charge freedom between the local paired states, and resulting in superconducting phase coherence in cuprates. Our experiment uncovers a new phenomenon in cuprate superconductors in which the intermediate states for phase coherence may exist, these states serve as the dynamical hopping of a hole between the local pairing states with two holes[54-56]. Our finding opens a new avenue to understand the origin of the superconductivity in cuprates, i.e., through local pairing and dynamical hopping of holes between the 'two-hole' bound states.

## Methods

**Sample preparation and characterization.** High-quality single crystals of $Bi_2Sr_{2-x}La_xCuO_{6+\delta}$ ($x$ = 0.75) were grown by using the traveling-solvent floating-zone technique[57]. Temperature-dependent resistivity measurements were carried out with a physical property measurement system (PPMS-9T,

Quantum Design). The DC magnetization measurements were performed with a SQUID-VSM-7T (Quantum Design).

**STM/STS measurements.** The STM/STS are performed with a cryogenic ultrahigh vacuum STM (USM-1300, Unisoku Co., Ltd.) with high magnetic field. Single crystals were cleaved at room temperature in ultrahigh-vacuum conditions up to $1.2 \times 10^{-10}$ torr. Sharp electrochemically etched tungsten tip was used in this study, and the calibration was carried out on the single crystalline Au (111) surface. The tunneling spectra were recorded by using a standard lock-in amplifier technique with an ac oscillation of 973 Hz. All the tunneling data in main text were taken at a temperature of about 1.5 K.

# Data availability

Relevant data supporting the key findings of this study are available within the article and the Supplementary Information file. All raw data generated during the current study are available from the corresponding author upon request.

# Acknowledgements

We thank Z.Y. Weng for helpful discussions, and we also acknowledge H. Q. Luo for the efforts in growing the single crystals. The work was supported by the National Key R&D Program of China (Grants No. 2022YFA1403201 and No. 2024YFA1408104), and the National Natural Science Foundation of China (Grants No. 11927809 and No. 12434004).

# Author contributions

The STM measurements were performed by H.L. under the supervision of H.Y. and H.H.W. H.L., Z.W., S.F. and J.X. analyzed the data and measured transport and magnetization properties of samples. H.H.W., H.Y. and H.L. wrote the paper. H.H.W. coordinated the whole work. All authors have discussed the results and the interpretations.

# Competing interests

The authors declare no competing interests.

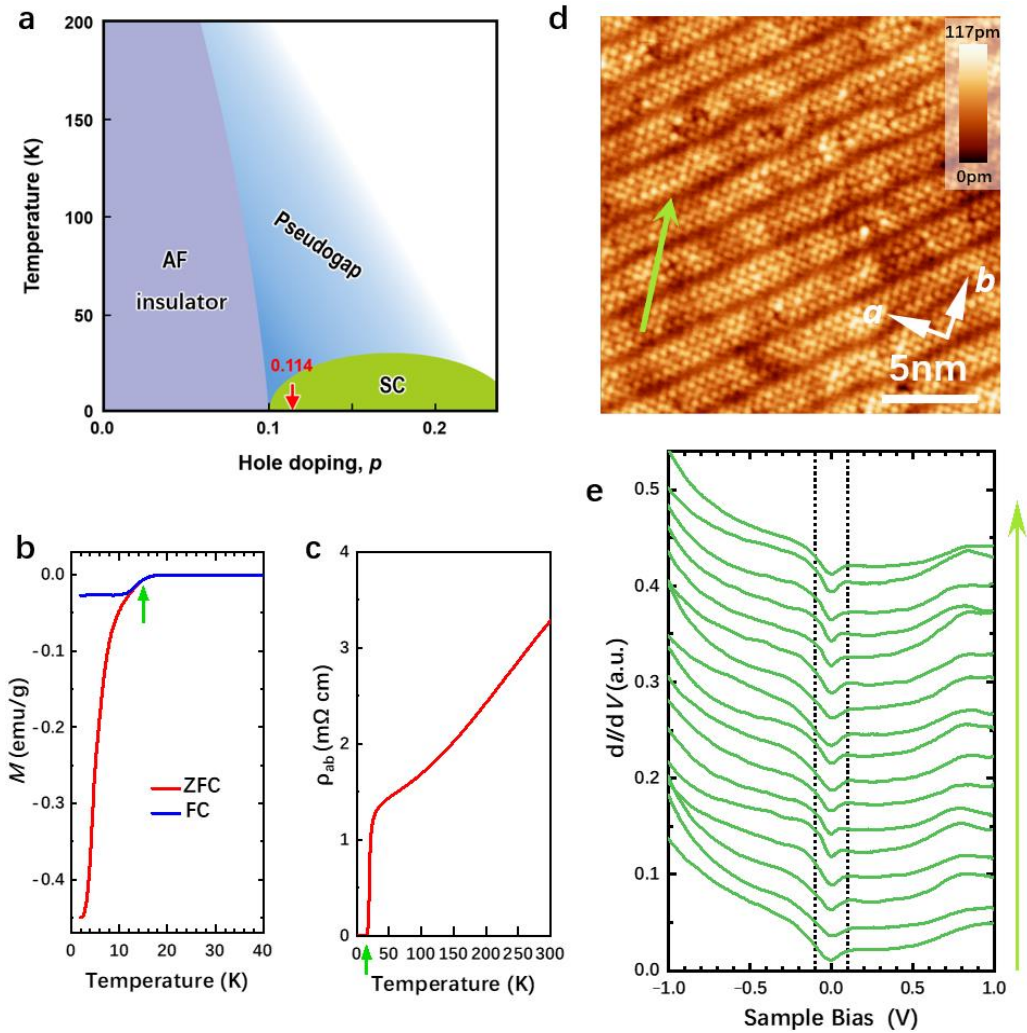

**Fig. 1| Phase diagram and characterization of the La-Bi2201 single crystals**. **a**, Schematic doping phase diagram of La-Bi2201. The hole-doping level of the present sample ($x = 0.75$, $p = 0.114$) is indicated by a red arrow. **b,c**, Temperature-dependent magnetization (**b**) and resistivity (**c**) of the La–Bi2201 sample with $p = 0.114$. The green arrows indicate zero resistance transition temperature at about 15 K. **d**, Atomically resolved topographic image measured on the La-Bi2201 sample with $p = 0.114$. Setpoint condition: $V_{set} = -400$ mV, $I_{set} = 50$ pA. **e**, A series of tunneling spectra taken along the green arrow in (**d**). Setpoint condition: $V_{set} = -1.0$ V, $I_{set} = 100$ pA. They are shifted vertically for clarity. The black dotted lines show the gapped feature at about $\pm 100$ meV.

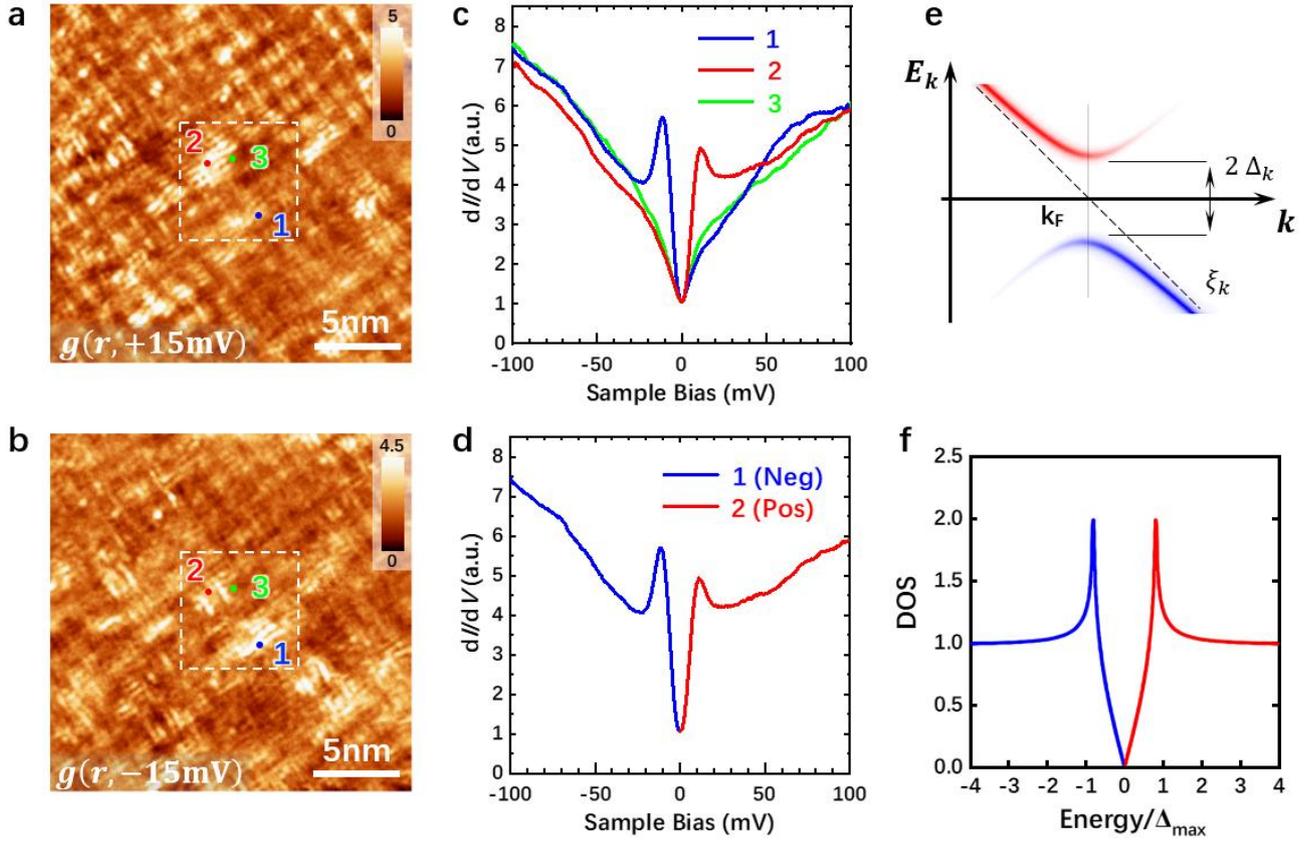

**Fig. 2| Electronic modulations and tunneling spectra with "half-Bogoliubon". a,b**, Differential conductance mappings measured at +15 and −15 mV, respectively. **c**, Typical tunneling spectra measured at different locations in **a** or **b**. Setpoint condition for **a-c**: $V_{set} = -250$ mV, $I_{set} = 100$ pA. The spectra are normalized by the differential conductance value at zero bias. Some of them show obvious particle-hole asymmetry, i.e., the coherence peak only exists on the positive or negative side, behaving as a feature of the "half-Bogoliubon". **d**, Artificially combined tunneling spectrum by merging together the negative-energy part of the spectrum measured at spot No. 1 and the positive part of the spectrum at spot No. 2. The combined spectrum behaves as a complete Bogoliubov dispersion for perfect superconductivity. **e**, Schematic image of the band dispersion in the normal (the dashed line) and superconducting (the red and blue curves) states of a conventional superconductor following the BCS theory. The blue and red curves indicate electron and hole branches of the energy band with

Bogoliubov quasiparticles, respectively. **f**, Ideal energy dependent DOS near Fermi energy of a *d*-wave superconductor.

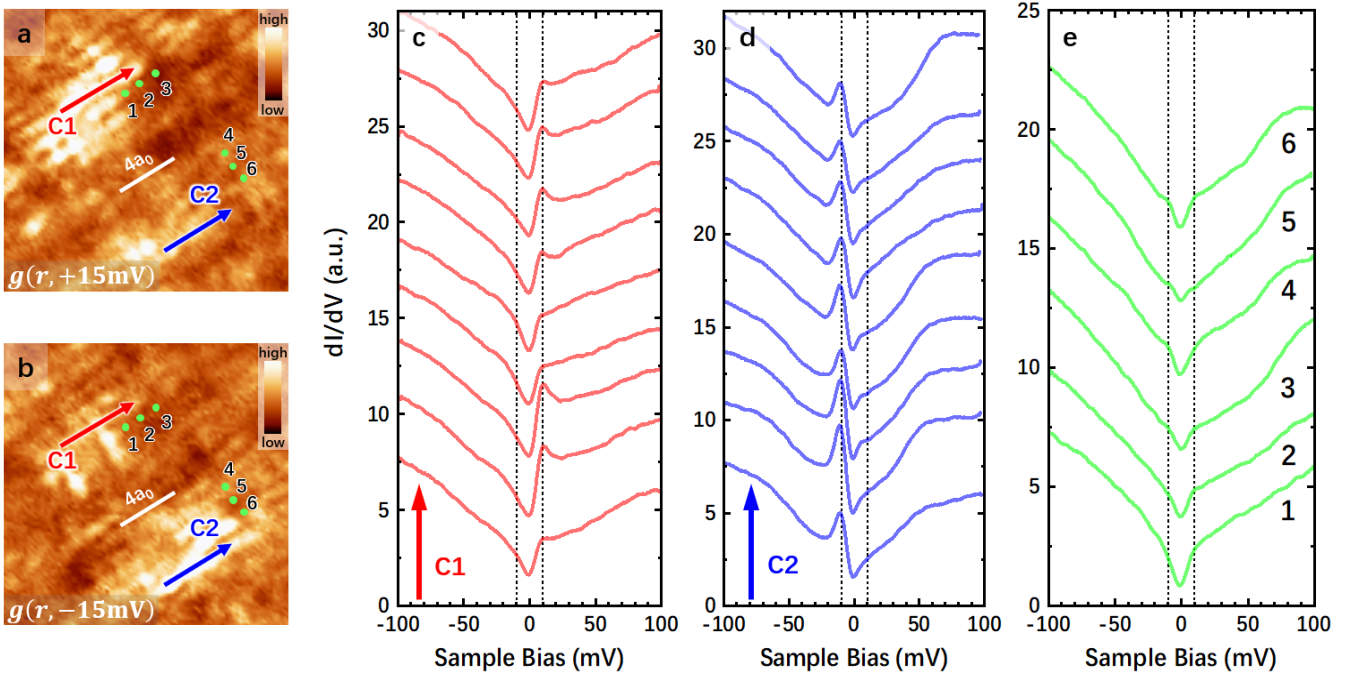

**Fig. 3| Particle-hole asymmetry of the unidirectional charge stripes. a,b**, Zoomed-in d$I$/d$V$ mappings within the white dashed square in Figs. 2a and 2b, respectively. The white lines indicate the length of 4$a_0$ **c**, Spatially resolved tunneling spectra measured along the arrowed line C1 in **a,b** on a stripe line bright at the positive bias. **d**, Spatially resolved tunneling spectra measured along the arrowed line C2 in **a,b** on a stripe line bright at the negative bias. **e**, Tunneling spectra measured on the background points from dots 1 to 6 in **a,b**. Tunneling spectra in **c-e** are shifted for clarity. Setpoint conditions: $V_{set} = -250$ mV and $I_{set} = 100$ pA.

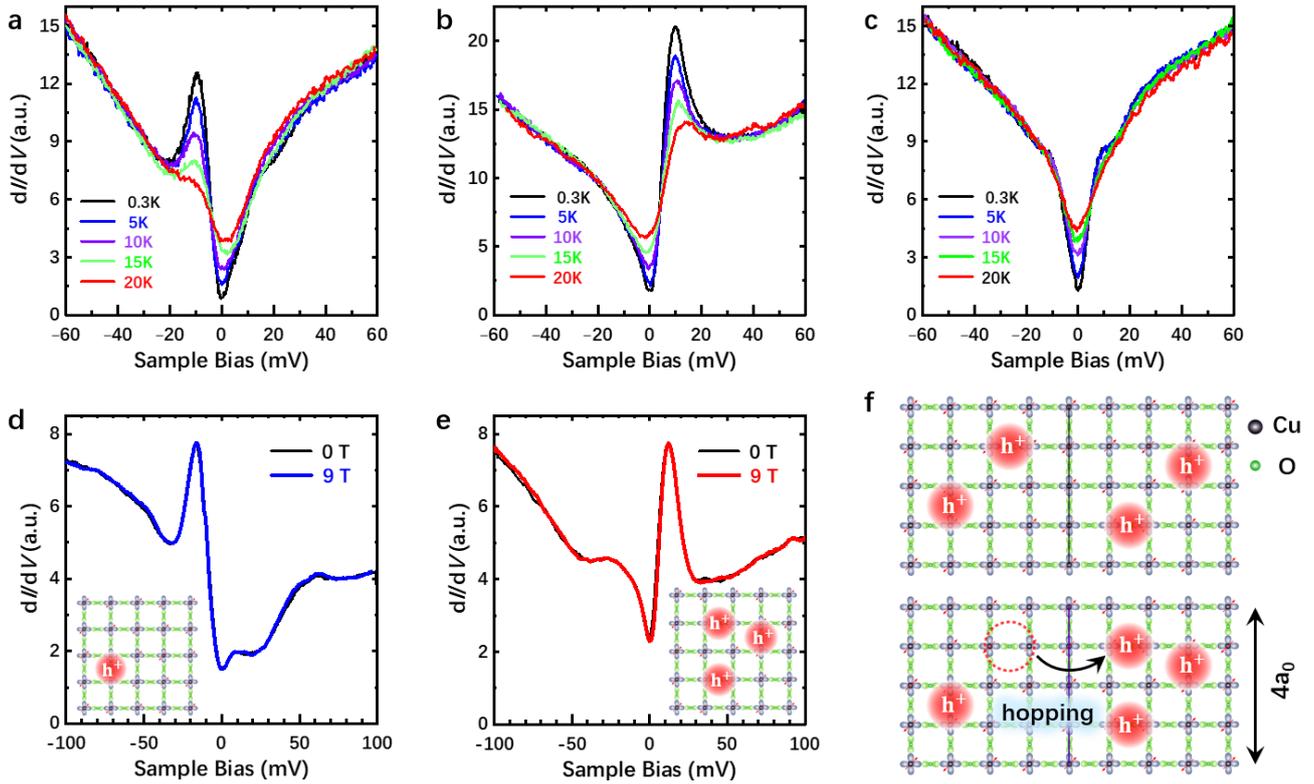

**Fig. 4| Temperature dependence and magnetic field insensitive evidences of the "half-Bogoliubons".** **a-c**, Tunnelling spectra taken at three different locations on the La-Bi2201 sample at various temperatures. **a-b**, The DOS of "half-Bogoliubons" at about +10 and −10 mV are both suppressed by increasing temperature. **c**, The spectrum with particle-hole symmetric gaps show a pair of kinks, this gap is clearly filled up by increasing temperature. **d,e** Tunnelling spectra measured under H = 0 T and 9 T magnetic field, respectively. It shows that neither positive nor negative energy "half-Bogoliubon" peaks are suppressed by a 9 T magnetic field. **f**, Schematic illustration of the dynamic hopping of charge freedom between local pairs of two holes. The upper panel shows the initial state that local paired 'two-hole' ground states within two adjacent $4a_0 \times 4a_0$ plaquettes. The lower panel demonstrates the breaken of a 'two-hole' paired state with one excited hole hopping into another plaquette. The temperature dependent spectra are taken by setup condition $V_{set}$ = −60 mV and $I_{set}$ = 100 pA, and the spectra taken under magnetic fields with $V_{set}$ = −250 mV and $I_{set}$ = 100 pA.

# Supplementary Materials for

# "Half – Bogoliubons" as the intermediate states for the phase coherence in underdoped cuprates

Han Li, Zhaohui Wang, Shengtai Fan, Jiaseng Xu, Huan Yang[✉], and Hai-Hu Wen[✉]

Center for Superconducting Physics and Materials, National Laboratory of Solid State Microstructures and Department of Physics, Collaborative Innovation Center for Advanced Microstructures, Nanjing University, Nanjing 210093, China

[✉]e-mail: huanyang@nju.edu.cn; hhwen@nju.edu.cn

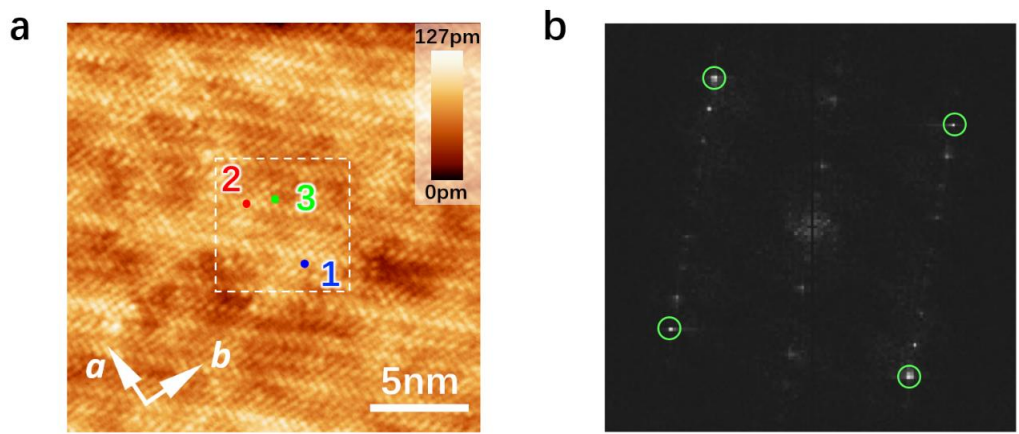

**Fig. S1: Topography and corresponding Fourier transform pattern. a**, The topography in the same field of view of the conductance mappings shown in Fig. 2. **b**, Fourier transformation of the topography shown in **a**, with green circles indicate the Bragg-peak positions.

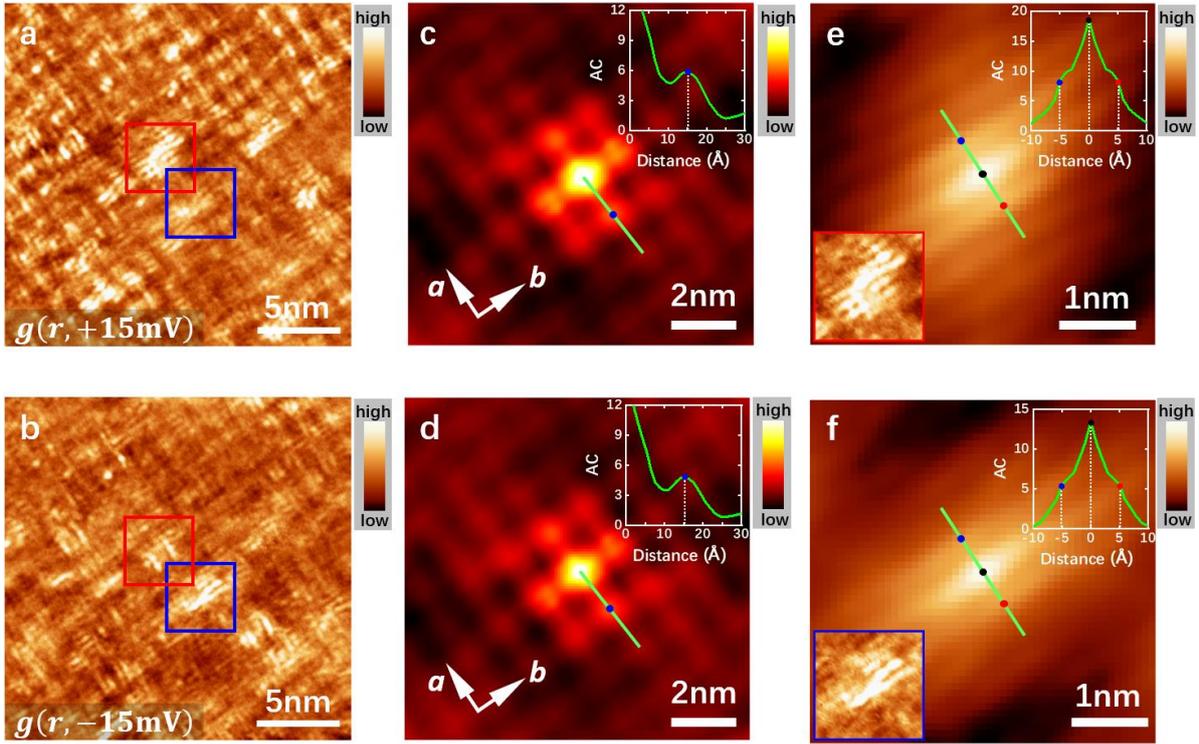

**Fig. S2: Charge stripes appearing on the background of $4a_0 \times 4a_0$ plaquette. a** and **b** are the conductance maps measured at 15 and -15 meV as shown in Fig.2, the spatially adjacent charge stripes at different bias are marked by the 4×4 nm² red and blue boxes, respectively. **c,** The autocorrelation map of **a** that revealing a long range $4a_0$-period plaquette. The linecut of autocorrelation intensity along Cu-Cu direction indicated by the green line is showing in the inset, blue dot reveal that the distance between neighboring charge puddles is about 15.2 Å. **d,** Autocorrelation map of **a** after substitute the value inside blue and red boxs with the average value of full map of **a**. The inset shows that the distance between neighboring charge puddles along the green linecut path is still about $4a_0$. **e,** The autocorrelation map of the area inside the red box in **a**, with the positive energy charge stripes enlarged in the corner inset. **f,** Autocorrelation map of the area inside the blue box in **b**, the inset at lower corner zooms in the negative energy charge strips. The linecut along the green path in **e** and **f** reveal the same $4/3a_0$ interval stripe distance for river-like stripe that bright on both positive and nagetive bias d$I$/d$V$ mapping. White arrows **a** and **b** are indicating the Cu-O bond direction.

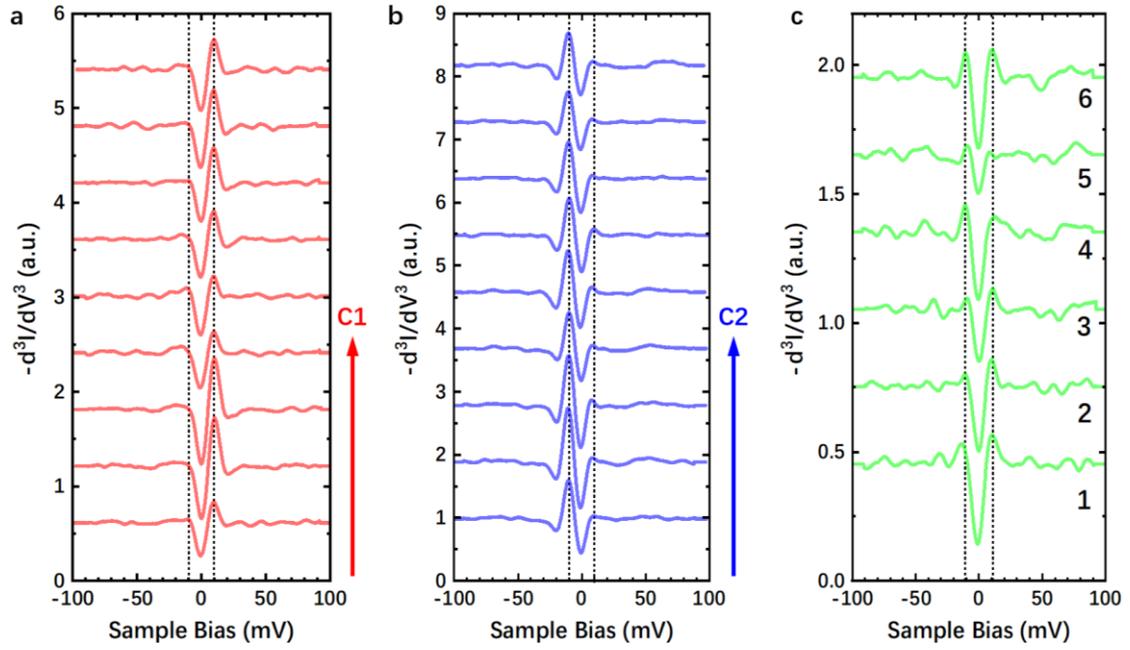

**Fig. S3: Highlight the peaks at ±11 mV by the second derative curves - of $d^3I/dV^3$ vs. V. a-c**, The $-d^3I/dV^3$ vs. V curves corresponding to the spectra taken along charge stripes C1, C2 and at points from dots 1 to 6 in Fig.3 **a,b**, respectively.

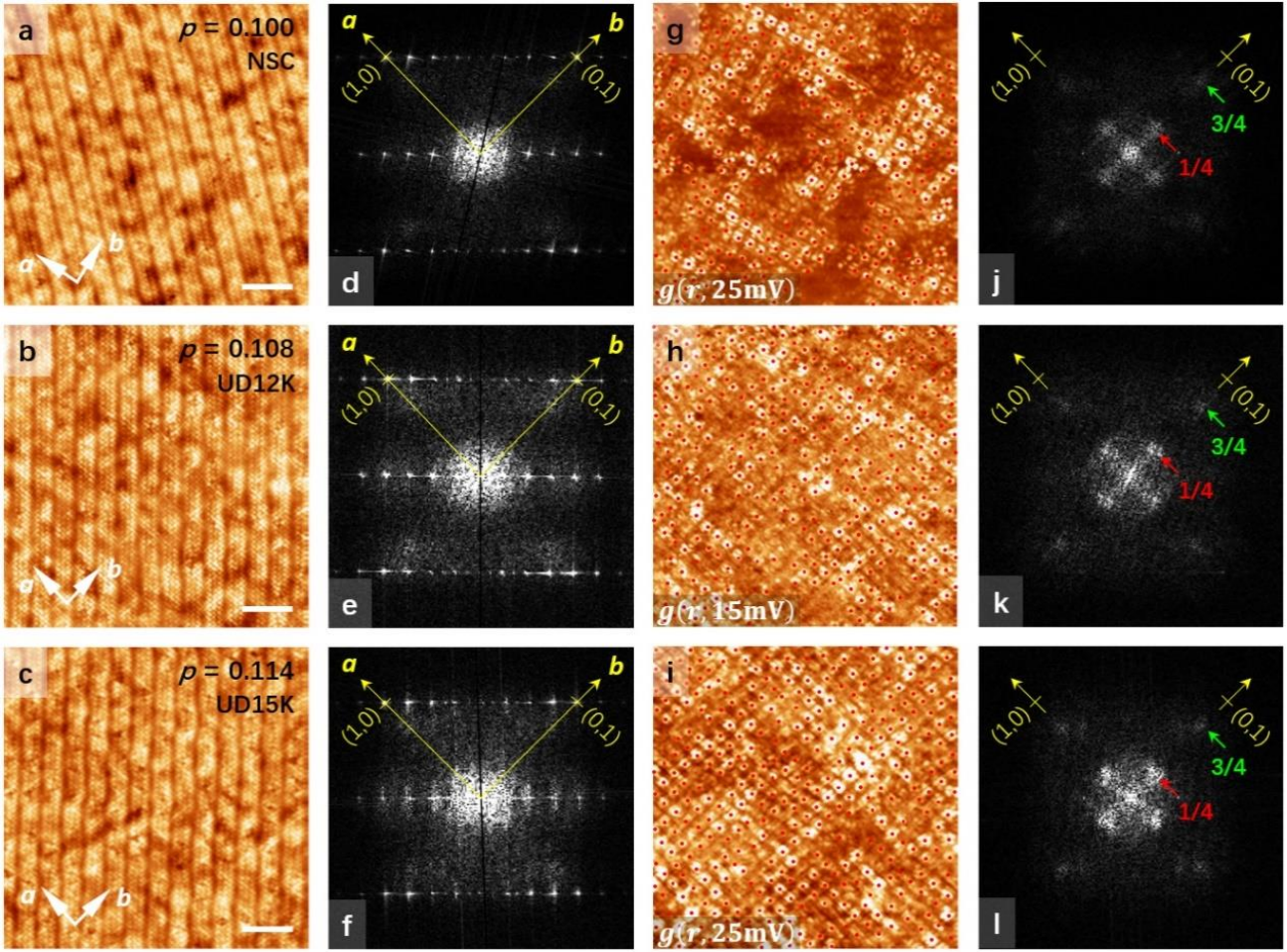

**Fig. S4: Doping dependent spatial occupancy of $4a_0$-period plaquettes. a-c**, Topographic images of three samples with nominal hole doping level $p$ = 0.100, 0.108, 0.114 in 30-nm$^2$ fields of view, and we denoted them as NSC, UD12 K and UD15 K, respectively. Here NSC represents nonsuperconducting as the $p$ = 0.100 sample is at the boundary between the insulating phase and the superconducting dome, UD represents underdoped and the number represents corresponding $T_c$. Orthogonal white arrows point the Cu-O bond directions marked by *a* and *b*, and scale bars indicate 5 nm. **d-f**, Fourier transformations of **a-c** are rotated for clear visualization, the Bragg peaks q = (±1, 0); (0, ±1)$2\pi/a_0$ correspongding to *a* and *b* directions are indicated by yellow crosses. **g-i**, Differential conductance mappings measured at the fields of view in **a-c**. The red dots denote the $4a_0$-plaquette locations and the number of plaquettes count 304, 332 and 346, given the estimated plaquette density

69.1%, 75.45% and 78.6%, respectively. If we assume that two holes occupied one plaquette averagely, then the hole density per Cu estimated for three samples are about $p = 0.098, 0.107$ and $0.112$, which are quite close to nominal doping level. **j-i**, Rotated Fourier transformations of **g-i**, in which the red and green arrows denote the $4a_0$ and $4a_0/3$ components. The data of $p = 0.10$ sample is taken by Cr tip at 2K, setup condition $V_{set} = -60$ mV and $I_{set} = 100$ pA. The data of $p = 0.108$ and $0.114$ samples are both taken by W tip at about 1.5K, setup $V_{set} = -200$ and $-250$ mV, $I_{set} = 100$ pA.